# Topological Thermal Hall Effect of Magnons in Magnetic Skyrmion Lattice

**Authors**

Masatoshi Akazawa[1]†, Hyun-Yong Lee[2,3,4]†, Hikaru Takeda[1], Yuri Fujima[5], Yusuke Tokunaga[5], Taka-hisa Arima[5], Jung Hoon Han[6]*, Minoru Yamashita[1]*

**Affiliations**

[1]The Institute for Solid State Physics, The University of Tokyo, Kashiwa, 277-8581, Japan

[2]Department of Applied Physics, Graduate School, Korea University, Sejong 30019, Korea

[3]Division of Display and Semiconductor Physics, Korea University, Sejong, 30019, Korea

[4]Interdisciplinary Program in E·ICT-Culture-Sports Convergence, Korea University, Sejong 30019, Korea

[5]Department of Advanced Materials Science, University of Tokyo, Kashiwa 277-8561, Japan

[6]Department of Physics, Sungkyunkwan University, Suwon 16419, Korea

(† These authors contributed equally to this work.)

**Abstract**

Topological transports of fermions are governed by the Chern numbers of the energy bands lying below the Fermi energy. For bosons, e.g. phonons and magnons in a crystal, topological transport is dominated by the Chern number of the lowest energy band when the band gap is comparable to the thermal energy. Here, we demonstrate the presence of topological transport by bosonic magnons in a lattice of magnetic skyrmions – topological defects formed by a vortex-like texture of spins. We find a distinct thermal Hall signal in the magnetic skyrmion phase of an insulating polar magnet $GaV_4Se_8$, identified as the topological thermal Hall effect of magnons governed by the Chern number of the lowest energy band of the magnons in a triangular lattice of magnetic skyrmions. Our findings lay a foundation for studying topological phenomena of other bosonic excitations through thermal Hall probe.

# Introduction

Non-trivial band topology of quasiparticles in a crystal realizes unique transport phenomena protected by the topology. In anomalous quantum Hall effect [1,2], a celebrated example of such topological transport predicted by Haldane, the dissipationless quantized current of conduction electrons is realized even in zero field owing to the topology of the electrons characterized by the finite Chern number of the energy bands. In these topological transports of fermions, such as electrons in insulators [1] and Majorana fermions in a Kitaev magnet [3], they occupy all the energy states below the Fermi energy, giving rise to a quantized current depending on the sum of the Chern numbers of the occupied bands.

Bosonic excitations in crystals can also exhibit topological transport when their energy bands acquire non-trivial topology. For example, magnons – collective excitations of spins – acquire non-trivial topology by the Dzyaloshinskii-Moriya (DM) interaction in ferromagnets [4]. The topological transport of charge-neutral excitations in insulators produces a thermal Hall effect (THE), a thermal analogue of the topological Hall effects by electrons. In an insulator, owing to the absence of the Hall effect by conduction electrons, the thermal Hall conductivity $\kappa_{xy}$ is simply given by

$$\frac{\kappa_{xy}}{T} = \frac{k_B^2}{\hbar} \int \Omega(E) f(E) dE, \qquad (1)$$

where $\Omega(E)$ and $f(E)$ are the Berry curvature and a distribution function given by the quantum statistics of the elemental excitations at the energy $E$, respectively [5,6]. Thus, $\kappa_{xy}/T$ in an insulator directly reflects the Berry-phase effect on the heat carriers as well as the quantum statistics that the heat carriers obey [7]. THEs in insulators have been observed in various magnetic insulators including ferromagnets [8], kagomé antiferromagnets [9–12], spin ices [13] and Kitaev materials [3,14–17].

In contrast to fermions, two or more bosons are allowed to occupy the same quantum state at low energies. Therefore, unlike fermions, the topological transport of bosons is expected to be strongly governed by the Chern number of the lowest energy band. A prominent example of such topological transport of bosons is the topological magnon Hall effect by magnetic skyrmions [18,19]. Each magnetic skyrmion produces an emergent magnetic field of order of 10–500 T inside the core, giving rise to a topological Hall effect for conduction electrons in metals [20–23]. This emergent field is also theoretically suggested to affect magnons and to realize a topological THE of magnons in which thermal currents carried by magnons are deflected by the magnetic skyrmions [24–29]. However, magnetic skyrmions in bulk materials are typically stable only near the magnetic ordering temperature, disabling the measurements down to low temperatures to observe the topological contribution from the lowest-energy magnon band.

Here, we demonstrate a topological THE of magnons realized by magnetic skyrmions in the polar magnet GaV$_4$Se$_8$ (Figs. 1(a)–(c)). The polar structure stabilizes cycloidal spin order with the magnetic modulation vectors perpendicular to the polar axis, realizing a Néel-type skyrmion phase down to the lowest temperature by applying a magnetic field parallel to the polar axis [30–33]. This stable skyrmion phase allows us to measure $\kappa_{xy}$ down to 0.2 K for investigating the effect of the lowest-energy magnon band. We find that a distinct thermal Hall conductivity emerges only in the skyrmion lattice phase but not in the forced-ferromagnetic nor in the cycloidal phase. Remarkably, $\kappa_{xy}$ does not depend on the field strength in the skyrmion phase, suggesting the topological nature of the THE. By theoretical calculations of the topological thermal Hal effect of the magnons in the triangular Néel-type skyrmion lattice, we indeed confirm that $\kappa_{xy}$ is determined by the Chern number of the lowest energy band. We further find the enhancement of $\kappa_{xy}$ by a magnetic-field poling and the difference in $\kappa_{xy}$ at low temperatures by entering the magnetic skyrmion phase from the low-field cycloidal and from the forced-ferromagnetic phase, indicating a high sensitivity of $\kappa_{xy}$ to the skyrmion dynamics.

## **Materials and Methods**

Single crystals of GaV$_4$Se$_8$ were synthesized by the chemical vapor transport method as described in the previous paper [30]. Magnetization was measured by using a magnetic property measurement system (MPMS, Quantum Design) under a magnetic field applied along the [111] axis. The thermal-transport measurements were performed by the steady-state method as described in Refs. [9–12] by using a variable temperature insert (VTI, 2–60 K, 0–15 T) and in a dilution refrigerator (DR, 0.15–4 K, 0–14 T). One heater and three thermometers were attached to the sample by using a silver paste, and then the temperature difference $\Delta T_x$ ( $\Delta T_x = T_\mathrm{H} - T_\mathrm{L1}$ ) and $\Delta T_y$ ( $\Delta T_y = T_\mathrm{L1} - T_\mathrm{L2}$ ) were measured as a function of the heat current $Q$ (see Fig. 1(c) for the configuration). To cancel the longitudinal component in $\Delta T_y$ by the misalignment effect, $\Delta T_y$ is asymmetrized with respect to the field direction as $\Delta T_y^{asym} = \Delta T_y(+B) - \Delta T_y(-B)$. The thermal (Hall) conductivity $\kappa_{xx}$ ($\kappa_{xy}$) is derived by

$$\begin{pmatrix} Q/wt \\ 0 \end{pmatrix} = \begin{pmatrix} \kappa_{xx} & \kappa_{xy} \\ -\kappa_{xy} & \kappa_{xx} \end{pmatrix} \begin{pmatrix} \Delta T_x/L \\ \Delta T_y^{asym}/w' \end{pmatrix}, \qquad (2)$$

where $t$ is the thickness of the sample, $L$ is the length between $T_H$ and $T_{L1}$, and $w$ is the sample width between $T_{L1}$ and $T_{L2}$, and $w'$ is the length between $T_{L1}$ and $T_{L2}$. To avoid the background Hall signal coming from a metal, the sample was attached to the insulating LiF heat bath.

All the data shown in Figs. 1, 2, 4 and 5 was obtained in sample 1. The magnetic-field poling measurements (Fig. 3) were done in sample 2. The standard error of all the data is smaller than the symbol size.

## Results
**Thermal-transport and magnetization measurements**

The lacunar spinel compound GaV$_4$Se$_8$ (Figs. 1(a) and 1(b)), a magnetic insulator in which each (V$_4$Se$_4$)$^{5+}$ cluster carries $S = 1/2$ spin, undergoes a structural transition from the non-centrosymmetric cubic to a polar phase below $T_S = 41$ K by elongating one of the <111> axes [30]. This first-order transition is clearly seen as a jump in the temperature dependence of the longitudinal thermal conductivity $\kappa_{xx}$ (Fig. 1(e)). A ferromagnetic interaction in the polar phase induces magnetic order below $T_C = 18$ K (Fig. 1(d)). The temperature dependence of $\kappa_{xx}$ shows a shoulderlike structure at ~22 K, which is followed by a peak just below $T_C$ (see arrows in Fig. 1(e)). The former shows the phonon contribution, whereas the latter shows a magnon contribution. This magnon contribution is also observed as the decrease in the field dependence of $\kappa_{xx}$ above the saturation field $B_{sat} \sim 0.4$ T (Fig. 2(c), see also Fig. S10 in Supplementary Materials (SM)). A competition between the ferromagnetic interaction and the DM interaction results in multiple non-collinear magnetic structures in GaV$_4$Se$_8$ below $T_C$. Previous magnetization [30] and elastic neutron scattering [31] studies have shown that a cycloidal phase at zero field turns into a Néel-type skyrmion phase above $B_{sky} \sim 0.1$ T applied along the [111] axis, which is followed by a transition to the forced-ferromagnetic phase above $B_{sat}$. One should note that, since the inversion symmetry is already broken above $T_S$, the structural transition results in four crystallographic polar domains with the polarization direction along the four cubic body diagonals, as observed in GaV$_4$S$_8$ [32–34]. The magnetic skyrmion phase appears in one of them in which the elongating axis is parallel to the magnetic field. In the other three domains, the cycloidal order is modified to conical order under a magnetic field, which is followed by a saturation above $B^* \sim$ 0.15 T. Each polar domain has a magnetic uniaxial anisotropy along the elongating axis [31], resulting in the gradual increase of $M$ above the saturation fields [32,33] (see

Fig. S1 in SM).

The field dependence of the magnetization $M$, $[\kappa_{xx}(B) - \kappa_{xx}(0)]/\kappa_{xx}(0)$, and $\kappa_{xy}/T$ at different temperatures are shown in Fig. 2 (see Methods and S2 in SM). In the paramagnetic phase above $T_C$ (Fig. 2(a)), $\kappa_{xx}$ monotonously increases with $B$, showing a typical field effect on phonons by suppressing spin fluctuations. A finite $\kappa_{xy}/T$ is observed only at high fields (Fig. 4(d)), which is vanishingly small below $B_{sat}$ (Fig. 2(a)). Below $T_C$, the magnetic transitions between the multiple magnetic phases are observed by the features in $\partial M/\partial B$ at $B_{sky}$, $B^*$, and $B_{sat}$ (top panels in Fig. 2) as reported in the previous study [30]. Although $\kappa_{xy}/T$ is absent just below $T_C$ (Fig. 2(b)), distinct $\kappa_{xy}/T$ with an almost flat field dependence is observed at 10 K in the magnetic skyrmion phase (Fig. 2(c)). In contrast, $\kappa_{xy}/T$ sharply disappears both in the cycloidal and forced-ferromagnetic phases. Moreover, the effect of the other domains at $B^*$ is not seen in $\kappa_{xy}/T$, which is in stark contrast to the field dependence of $\kappa_{xx}$ changing intricately at all magnetic transitions. These results demonstrate the high sensitivity of $\kappa_{xy}$ to the emergence of the magnetic skyrmion phase. As shown in Figs. 1(f) and 1(g), the color plot of $\kappa_{xy}/T$ in the $B$–$T$ phase diagram clearly indicates the stable region of the magnetic skyrmion phase. The reproducibility of $\kappa_{xy}$ in the magnetic skyrmion phase is confirmed in the measurements of another sample (see S3 in SM), as well as the measurements done in different cryostats (Figs. 2(a)–(e) and Fig. 2(f), see S4 in SM).

**Magnetic-field poling effect**
To further confirm the field dependence both in $\kappa_{xx}$ and $\kappa_{xy}$ in the magnetic skyrmion phase, we investigate the magnetic-field poling effects in sample 2 by cooling through $T_S$ under a finite magnetic field (7 T and 15 T for the magnetization and the thermal-transport measurements, respectively. See S5 in SM for more details). Because of the magnetic anisotropy, the domain volume of the magnetic skyrmion phase (***M*** ∥ ***B***) is expected to be increased by applying a magnetic field at $T_S$, as observed in the related compound $GaV_4S_8$ (Ref. [34]). In the field dependence of $M$, the magnetic skyrmion domain shows a step-like increase of $M$ both at $B_{sky}$ and $B_{sat}$. On the other hand, $M$ of the conical phase in the other domains increases in proportional to $B$, which is followed by a saturation above $B^*$. In fact, as shown in Fig. 3(a), $M$ measured after the magnetic-field poling (MP) shows (1) a smaller slope of $M$ below $B^*$, (2) sharper increases of $M$ at $B_{sky}$ and $B_{sat}$, and (3) the increase of $M$ above $B_{sat}$, compared to $M$ measured after zero magnetic-field poling (ZMP). These results confirm that the domain volume of the magnetic skyrmion phase is increased by MP.

We find that, whereas this MP does not affect $\kappa_{xx}$ at zero field (Fig. S9), it drastically changes the field dependence of both $\kappa_{xx}$ and $\kappa_{xy}$ (Fig. S10). First, the gradual increase of $\kappa_{xx}$ observed in ZMP in the field range of $0 < B < B^*$ disappears by MP and a clear drop of $\kappa_{xx}$ is observed at $B_{\text{sky}}$ (Fig. 3(b)). The gradual increase of $\kappa_{xx}$ in the ZMP is brought by the increase of the phonon conductivity in the conical domain, in which the magnon-phonon scattering is decreased by asymptotical polarization of the conical phase into the forced-ferromagnetic phase. Therefore, the disappearance of the gradual increase of $\kappa_{xx}$ by MP also shows the increase of the skyrmion domain volume. The drop of $\kappa_{xx}$ at $B_{\text{sky}}$ in MP shows the backflow effects of magnetic skyrmions [25,26] and/or additional scattering effects on magnons and phonons by the magnetic skyrmions. Most remarkably, whereas $\kappa_{xy}$ in ZMP is similar both in the 1st and 2nd run, the magnitude of $\kappa_{xy}$ after MP becomes about three times larger than that measured in ZMP (Fig. 3(c)). All these results obtained by MP demonstrate the genuine thermal-transport features of the magnetic skyrmion phase (see S5 in SM for details).

**Hysteresis in the magnetic field**

As shown in Fig. 2, $B_{\text{sat}}$ shows a large hysteresis in the magnetizing ("$|B|$ up") and in the demagnetizing ("$|B|$ down") measurements. This hysteresis is caused by the energy barrier for nucleation/annihilation of the magnetic skyrmion, resulting in the first-order transition at $B_{\text{sat}}$. In addition to this magnetic hysteresis, we find a large hysteresis in the field dependence of $\kappa_{xy}/T$ emerged at lower temperatures (see Figs. 2(d)–(f)). When entering the magnetic skyrmion phase from the forced-ferromagnetic phase ($|B|$ down), $\kappa_{xy}/T$ becomes smaller below 5 K and is almost absent below 2 K. On the other hand, when entering the magnetic skyrmion phase from the cycloidal phase ($|B|$ up), the field dependence $\kappa_{xy}/T$ turns to be a shoulder shape with a peak near the boundary to the forced-ferromagnetic phase. The temperature dependence of $\kappa_{xy}/T$ in the magnetic skyrmion phase is plotted in Fig. 4(e). As shown in Fig. 4(e), $\kappa_{xy}/T$ in the $|B|$ down process shows a peak at around 6 K. On the other hand, $\kappa_{xy}/T$ in the $|B|$ up process increases down to 2 K, which is followed by a rapid decrease to zero at lower temperatures (Fig. 4(e)). The peak in the field dependence of $\kappa_{xy}/T$ in the $|B|$ up process is well resolved even at 0.2 K (Fig. 2(f)), showing the persistence of the magnetic skyrmion phase in GaV$_4$Se$_8$ down to the lowest temperature. As shown in Fig. 1(f) and Fig. 1(g), this large hysteresis in $\kappa_{xy}/T$ manifests as a different color profile of the $B$–$T$ phase diagram in the two processes. Moreover, $\kappa_{xy}/T$ in the $|B|$ up process is suppressed in the intermediate phase (Fig. 1(f) and Fig. S5), implying a different magnetic state in the intermediate phase [31].

## Discussion
**Origin of the thermal Hall effect in the magnetic skyrmion phase**

In a magnetic insulator such as GaV$_4$Se$_8$, one can think of four possible origins for the THE; a conventional magnon THE ($\kappa_{xy}^{\text{mag}}$), a topological magnon THE ($\kappa_{xy}^{\text{topo}}$), a skyrmion Hall effect ($\kappa_{xy}^{\text{skr}}$), and a phonon THE ($\kappa_{xy}^{\text{ph}}$). In $\kappa_{xy}^{\text{mag}}$, the magnons execute Hall-like transport by virtue of the "emergent magnetic field" originating from the DM interaction [5,8], in which the main role of the external magnetic field is to lift the degeneracy between the up and down spins [35]. By contrast, in $\kappa_{xy}^{\text{topo}}$, magnons are still deflected by emergent magnetic field, but this time the origin is the magnetic skyrmions [24,25,27–29]. The external magnetic field stabilizes the skyrmion phase while the magnons couple to the skyrmion spin texture in its dynamical motion. The skyrmion Hall effect $\kappa_{xy}^{\text{skr}}$ refers to the reduction in $\kappa_{xy}^{\text{topo}}$ due to the depinned motion of the skyrmion lattice relative to the magnon flow [36–39]. THEs of phonons have been observed in various materials [12,40–44], and theoretically studied [45–50]. Below, we discuss these four origins and argue that the THE observed in the magnetic skyrmion phase of GaV$_4$Se$_8$ is most likely of topological origin dominated by $\kappa_{xy}^{\text{topo}}$.

The conventional magnon THE occurs even in the ferromagnetic phase [8] and should, in principle, be observable in the forced-ferromagnetic phase. The small $\kappa_{xy}$ just above $B_{\text{sat}}$ (Fig. 2(d) and 2(e)) would thus give un upper limit of $\kappa_{xy}^{\text{mag}}$ in this compound. The field dependence of $\kappa_{xy}^{\text{mag}}$ is mainly given by a polylogarithmic function of $e^{-g\mu_B B/k_B T}$, where $g$ is the g-factor and $\mu_B$ the Bohr magneton [8]. We estimate this field dependence $\kappa_{xy}^{\text{mag}} \propto e^{-g\mu_B B/k_B T}$ as the gray solid line shown in Fig. 2 and in Fig. S4, indicating a negligible contribution of $\kappa_{xy}^{\text{mag}}$ except for the 2 K data in $|B|$ down process in which $\kappa_{xy}^{\text{topo}}$ already disappears as shown by our calculation discussed below (Fig. 5(e)). We can thus safely conclude that $\kappa_{xy}^{\text{mag}}$ is negligible in the magnetic skyrmion phase.

Skyrmion Hall effect only exists when the heat current exceeds the depinning threshold and the magnetic skyrmion lattice is set adrift by the magnon flow [26,37] (see S6 in SM for the heat current dependence studied in GaV$_4$Se$_8$), which has been observed for magnetic skyrmions in metals [36,38,39]. The transverse deflection of magnons against the (moving) skyrmions in turn creates a reaction by the magnons on the skyrmion, driving the latter object to the transverse direction of the overall drift and resulting in the reduction of the transverse heat flow (Fig. 5(a)). In the extreme limit of the average magnon velocity equaling the skyrmion drift velocity and no relative motion between the two, both the skyrmion Hall effect and topological magnon THE must vanish since neither the magnons nor the skyrmions will be exerting transverse force on the other. By its nature, the skyrmion Hall effect is never a dominant effect, but provides a reduction mechanism to the existing topological Hall effect. Further, the sign of $\kappa_{xy}^{\text{sky}}$ is shown to be negative by the simulation [26], which is opposite to our experimental data of $\kappa_{xy} > 0$ measured in the magnetic skyrmion phase.

The THEs of phonons $\kappa_{xy}^{\text{ph}}$ is expected to be linear in the magnetic field because the energy scale of the phonons given by the Debye temperature (several hundred Kelvin) is much larger than that of the magnetic fields (up to 15 T). In addition, $\kappa_{xy}^{\text{ph}}$ is known to scale well with $\kappa_{xx}$ and reach the peak at the same temperature as of $\kappa_{xx}$ as observed in various materials [12,43,44]. In GaV4Se8, we indeed find that $\kappa_{xy}$ measured up to 15 T shows an almost linear field dependence at high fields (Figs. 4(a)–(d)). Furthermore, the temperature dependence of $\kappa_{xy}/T$ at 15 T scales well to $\kappa_{xx}/T$ (black circles in Fig. 4(e)), showing a finite contribution of $\kappa_{xy}^{\text{ph}}$ at $B \gg B_{\text{sat}}$. However, extrapolating this $\kappa_{xy}^{\text{ph}}$ to $B < B_{\text{sat}}$ results in a negligible $\kappa_{xy}^{\text{ph}}$ in the magnetic skyrmion phase. We can also rule out the possibility that the phonon THE is somehow enhanced in the magnetic skyrmion phase by the large emergent field produced by the magnetic skyrmions. First, as shown in Fig. 4(e), $\kappa_{xy}/T$ in the magnetic skyrmion phase do not scale to that of $\kappa_{xx}/T$, in particular for $\kappa_{xy}/T$ in the $|B|$ up process, lacking the telltale sign of the phonon origin observed in other materials [12,43,44]. This mismatch of the peak temperature is in contrast to the good scaling observed for $\kappa_{xy}^{\text{ph}}/T$ at high fields and $\kappa_{xx}/T$ (Fig. 4(e)). Second, this topological $\kappa_{xy}^{\text{ph}}$ in the magnetic skyrmion phase is inconsistent with the difference in $\kappa_{xy}/T$ in the magnetizing and demagnetizing processes at low temperatures. As shown in the top panels in Fig. 2, $M$ does not show a marked difference in the two processes except that caused by the different phase boundary of the forced-ferromagnetic phase owing to the first-order transition nature, suggesting that the number of magnetic skyrmions remains nearly the same. In the topological $\kappa_{xy}^{\text{ph}}$ scenario, the amount of Hall transport will be proportional to the skyrmion density. The magnetization data indicates that the skyrmion density remains the same for magnetization and demagnetization processes, but the thermal Hall data exhibits sizable difference in the two cases. Also on theoretical grounds, it is extremely difficult to envision a mechanism by which to couple the phonon angular momentum to the magnetization texture. For these various reasons we find it safe to exclude the phonon origin of the observed THE.

From the considerations above, we conclude that the THE in GaV$_4$Se$_8$ is dominated by $\kappa_{xy}^{\text{topo}}$ as a result of the magnons seeing the emergent field of skyrmions. We further note that, whereas magnons do not significantly contribute to the longitudinal thermal transport (Fig. 1(e)), they can dominate the thermal Hall effect as observed in the ferromagnet Lu$_2$V$_2$O$_7$ [8].

**Simulation of topological thermal Hall effect of magnons in magnetic skyrmion lattice**

The dominating contribution of $\kappa_{xy}^{\text{topo}}$ in our data is further supported by our theoretical investigation of the temperature dependence of $\kappa_{xy}^{\text{topo}}$ arising from the magnon-skyrmion interaction. This interaction has been discussed in terms of the Berry curvature effects on the magnon band [4–7,24,28] formed in a lattice of magnetic skyrmions and the scatterings of magnons [25–27] by the individual skyrmions. We adopt the former momentum-based picture rather than the latter real-space-based one because we need to understand the magnon transport in a lattice of magnetic skyrmions formed in GaV$_4$Se$_8$ (see also S7 in SM).

We solve the magnon band problem in the background of a triangular lattice of Néel-type skyrmions (Fig. 5(a)). The local spin configuration of the skyrmion spin texture is given by $\vec{n}_i = (\sin\theta_i \cos\phi_i, \sin\theta_i \sin\phi_i, \cos\theta_i)$. It is convenient to perform the rotation of the spin quantization axis so that the new spin direction becomes aligned with $\vec{n}_0 = (0,0,1)$. The rotation of the frame gives rise to an emergent vector potential and gauge field to the magnon excitation leading to the topological magnon modes [19]. The emergent vector potential appears in the form of a phase factor for the bosons (magnons) and gives rise to a bosonic tight-binding problem

$$H = -SJ \sum_{\langle ij \rangle} \left( t_{ij} b_i^\dagger b_j + h.c. \right) \quad (3)$$

[$S$ = size of magnetic moment, $J$ = spin interaction energy, $b_i$ = boson amplitude at lattice site $i$]. Crucially, the hopping phase term $t_{ij}$ is affected by the local spin texture [19] according to $t_{ij} = e^{2i(\vec{m}_i \times \vec{m}_j) \cdot \vec{n}_0}$ where $\vec{m}_i$ is given locally by $\left( \sin \frac{\theta_i}{2} \cos \phi_i, \sin \frac{\theta_i}{2} \sin \phi_i, \cos \frac{\theta_i}{2} \right)$ and $\vec{n}_0 = (0,0,1)$.

One chooses $\vec{n}_i$ to be that of the triangular lattice of the Néel-type skyrmions (Fig. 5(a)) and diagonalize the bosonic Hamiltonian in Eq. (3) accordingly. The phase $2 \left( \vec{m}_i \times \vec{m}_j \right) \cdot \vec{n}_0$ implies the magnons accumulate the phase worthy of two flux quanta as they hop around a skyrmion [19]. A careful numerical simulation concluded that a skyrmion deflects the magnon and electron in the same direction [37], which led us to adopt the + sign for the magnon hopping phase. After the Fourier transformation into the momentum space, the magnon Hamiltonian is recast as

$$H = -SJ \sum_{\vec{k} \in BZ} \sum_{n=1}^{N_u} \sum_{m \in n} \left[ t_{nm} e^{i\vec{k} \cdot (\vec{r}_n - \vec{r}_m)} + c.c \right] b_{\vec{k}n}^\dagger b_{\vec{k}m}, \quad (4)$$

where $N_u$ stands for the number of sites in the unit-cell and $\vec{r}_n$ for the position of site $n$ in the magnetic unit-cell. By diagonalization of the $N_u \times N_u$-matrix, one acquires the magnon spectrum $\varepsilon_{n\vec{k}}$ for a given momentum $\vec{k}$. In our simulation, we employ a 64-site unit-cell of the Néel-type skyrmion lattices, and thus 64 magnon bands appear in the Brillouin zone. Note that the spectrum of the Hamiltonian in Eq. (4) needs not be positive-definite while a full consideration of the spin interactions and magnetic field should make it so. We therefore add a constant to the overall spectrum to guarantee the positive-definite spectrum with a tiny gap equal to 0.1% of the entire bandwidth.

The Berry curvature of the $n$-th band can be evaluated by the following formula [51,52]:

$$\Omega_n(\vec{k}) = -2 \sum_{m \neq n} \frac{\text{Im}[\langle n, \vec{k}|v_x|m, \vec{k}\rangle \langle m, \vec{k}|v_y|n, \vec{k}\rangle]}{(\varepsilon_{n\vec{k}} - \varepsilon_{m\vec{k}})^2}, \quad (5)$$

where $v_\gamma = \partial H / \partial k_\gamma$. The Berry curvature distribution of the lowest and second-lowest bands in the momentum space are shown in Fig. 5(b) and Fig. 5(c), respectively.

Integrating the Berry curvature over the Brillouin zone, one obtains the Chern number of the $n$-th band, $C_n = \frac{1}{2\pi} \int_{BZ} \Omega_n(\vec{k})$. The ten lowest magnon bands are shown in Fig. 5(d) along with the Chern number. Importantly, the lowest band turns out to be topological with a Chern number $C_1 = +1$ as specified in Fig. 5(d), while the second and third bands are trivial with zero Chern number (a full list of Chern numbers for the bands is given in Table 1).

We compute $\kappa_{xy}^{topo}$ by using Eq. (1) with

$$f(E) = -c_2 \left[ n_B \left( \frac{E}{k_B T} \right) \right], \tag{6}$$

where $c_2(x) = (1+x)\left(\ln\frac{1+x}{x}\right)^2 - (\ln x)^2 - 2\text{Li}_2(-x)$ , $n_B(x) = (e^x - 1)^{-1}$ the Bose-Einstein distribution function, and Li$_2$ is the polylogarithm function [6]. We adopt the mean-field-like temperature dependence of the magnetization, $S = \sqrt{1 - T/T_c}$, and set the critical temperature $T_c \sim J$ (the entire magnon band width is about $10J$). We present the result in Fig. 5(e) as a function of temperature. To facilitate comparison to the calculation, we have estimated $\kappa_{xy}^{topo}$ of the two-dimensional layer of the magnetic skyrmions ($\kappa_{xy}^{2D}$) from the experimental data at a fixed field by $\kappa_{xy}^{2D} = \kappa_{xy} d$, where $d = 0.5854$ nm is the interlayer distance (Fig. 1(a)). Given that the skyrmion phase appears only in one of the four domains, we further multiply $\kappa_{xy}^{2D}$ by a constant in Fig. 5(e) (S5 in SM).

**Comparison between the $\kappa_{xy}$ measurements and the simulation**
As the temperature increases, the magnon bandwidth, governed by the size of the ordered moment, decreases while the thermal factor $k_B T$ allows more of the higher-energy bands to contribute to Hall transport. The sum of all Chern numbers is zero, $\sum_n C_n = 0$, and consequently $\kappa_{xy}^{topo}$ must vanish at a sufficiently high temperature. The calculation shows a nearly complete suppression of $\kappa_{xy}^{topo}$ at temperatures above $0.7T_C$ and a peak at about $0.2T_C$. These good agreements with the experimental result (Fig. 5(e)) strengthen our topological magnon interpretation of the observed $\kappa_{xy}$. Moreover, $\kappa_{xy}^{topo}$ peaks around the temperature comparable to the energy of the lowest magnon band with high $\Omega$ ("high-$\Omega$ spot" in Fig. 5(d)), providing support that topological magnons from the lowest band dominate the low-temperature $\kappa_{xy}^{topo}$. We present the Berry curvature of the lowest and the second lowest bands in Fig. 5(b) and Fig. 5(c), respectively. It is positive and peaked at $K$ point in the lowest band (Fig. 5(b)), while negative-valued peaks appear at $\Gamma$ and $K$ points in the second band (Fig. 5(c)). This positive Berry curvature is given by the topological charge of the magnetic skyrmion itself. Whereas the magnon itself is charge neutral, it couples to the emergent magnetic field of the skyrmions effectively as a charge +1 quasiparticle. As $k_B T$ exceeds the peak temperature, the negative Berry curvature contribution from the second magnon band begins to undermine the lowest-band contribution to $\kappa_{xy}^{\text{topo}}$.

The good agreement between the $\kappa_{xy}$ data and our simulation of topological THE of magnons indicates that the observed THE is indeed a topological magnon THE. We reveal that magnetic skyrmions do affect the magnonic heat carriers in magnetic insulators as they affect conduction electrons in magnetic metals. We note that, while $\kappa_{xy}/T$ observed in the $|B|$ down process agrees much better to the theoretical simulation, the peak of $\kappa_{xy}/T$ occurs at a clearly lower temperature in the $|B|$ up process (Fig. 5(e)). Given that the peak temperature of $\kappa_{xy}/T$ roughly corresponds to the energy level of the high-$\Omega$ spot, the lower peak temperature of $\kappa_{xy}/T$ in the $|B|$ up process implies that the energy level of the high-$\Omega$ spot is lowered for some reason. It should be noted that a possibility that the decrease of $\kappa_{xy}/T$ below 2 K is caused by another magnetic transition within the skyrmion phase can be safely ruled out because of the absence of an anomaly in both the temperature dependence of $\kappa_{xx}$ and the field dependence of $\kappa_{xy}$ below 2 K (see section S4 in SM). Meanwhile, the small difference in $M$ in the two processes (top panels of Fig. 2) shows that almost the same number of magnetic skyrmions are created in the two processes. Therefore, we conclude the difference in $\kappa_{xy}$ is most likely due to a subtle rearrangement in the skyrmion lattice structure between the two magnetization processes which in turn affect the magnon band structure. Although this possibility should be scrutinized by direct measurements of the hysteresis effect on the skyrmion lattice in the future, $\kappa_{xy}$ may provide an unusually sensitive probe of the skyrmion lattice structure not attainable by other measurements.

Note that other models for magnon bands in the presence of skyrmion texture exist and give different Chern number distributions of the magnon bands [28,53,54] from ours. The resulting $\kappa_{xy}$ and its temperature dependence could be dissimilar to ours. We leave this question for future investigation. Regardless of model details, a general lesson remains that the temperature dependence of the thermal Hall conductivity serves as a sensitive probe of the Berry curvature and the energy distribution of the magnon bands.

**Acknowledgements:**

We thank Kouki Nakata for fruitful discussions. This work was supported by Grants-in-Aid for Scientific Research (KAKENHI) (No. JP19H01848, No. JP19K21842, and No. JP19H05826). H.-Y.L. was supported by a Korea University Grant and National Research Foundation of Korea (NRF- 2020R1I1A3074769). Y.F. was supported by the Japan Society for the Promotion of Science through the Program for Leading Graduate Schools (MERIT) and the JSPS Research Fellowship for Young Scientists (JSPS KAKENHI Grant No. JP18J13415).

# Figures and legends:

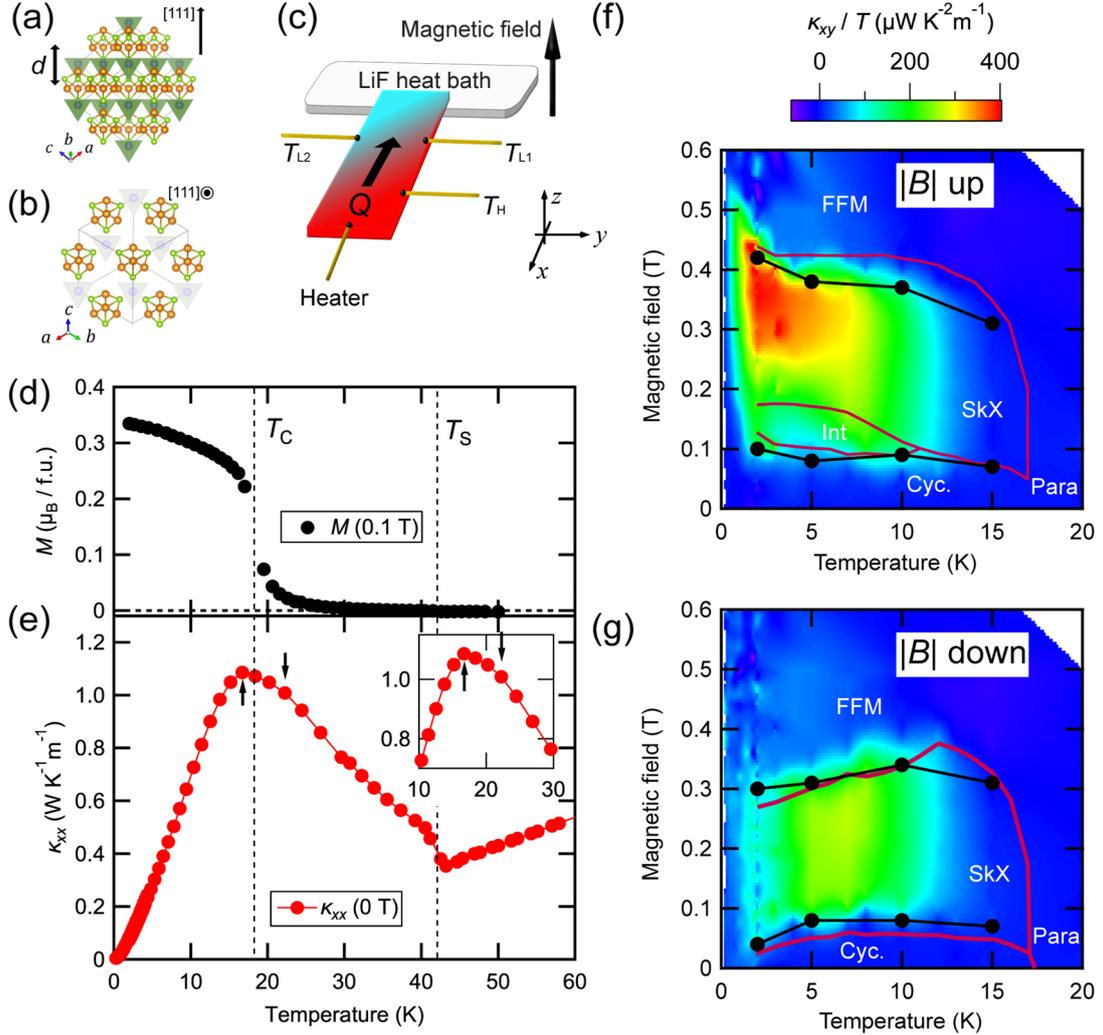

**Fig. 1 Magnetic skyrmion phase in GaV$_4$Se$_8$.** (a, b) Crystal structure of GaV$_4$Se$_8$ viewed parallel (a) and perpendicular (b) to the plane of (V$_4$Se$_4$)$^{5+}$ clusters drawn by VESTA software [55]. The double arrow in (a) shows the inter-layer distance $d$. (c) An illustration of the experimental setup. One heater and three thermometers ($T_H$, $T_{L1}$, and $T_{L2}$) are attached to detect both the longitudinal and the transverse temperature gradients (shown by the exaggerated color gradation for a clarity) in the sample fixed on the LiF heat bath (see Methods). (d, e) The temperature dependence of the magnetization ($M$, d) at 0.1 T and the longitudinal thermal conductivity ($\kappa_{xx}$, e) at zero field. The structural transition temperature ($T_S$) and the magnetic ordering temperature ($T_C$) are shown. The inset shows an enlarged view near the two features indicated by arrows. (f, g) Color plots of the thermal Hall conductivity ($\kappa_{xy}/T$) measured in the magnetizing (|$B$| up, f) and demagnetizing (|$B$| down, g) procedure. Paramagnetic (Para), cycloidal (Cyc), magnetic skyrmion (SkX), intermediate (Int) [31], and forced-ferromagnetic (FFM)

phases are indicated. The red solid lines and black circles show the magnetic phase boundary determined by the previous measurements [30,31] and by the magnetization measurements (top panels of Fig. 2), respectively.

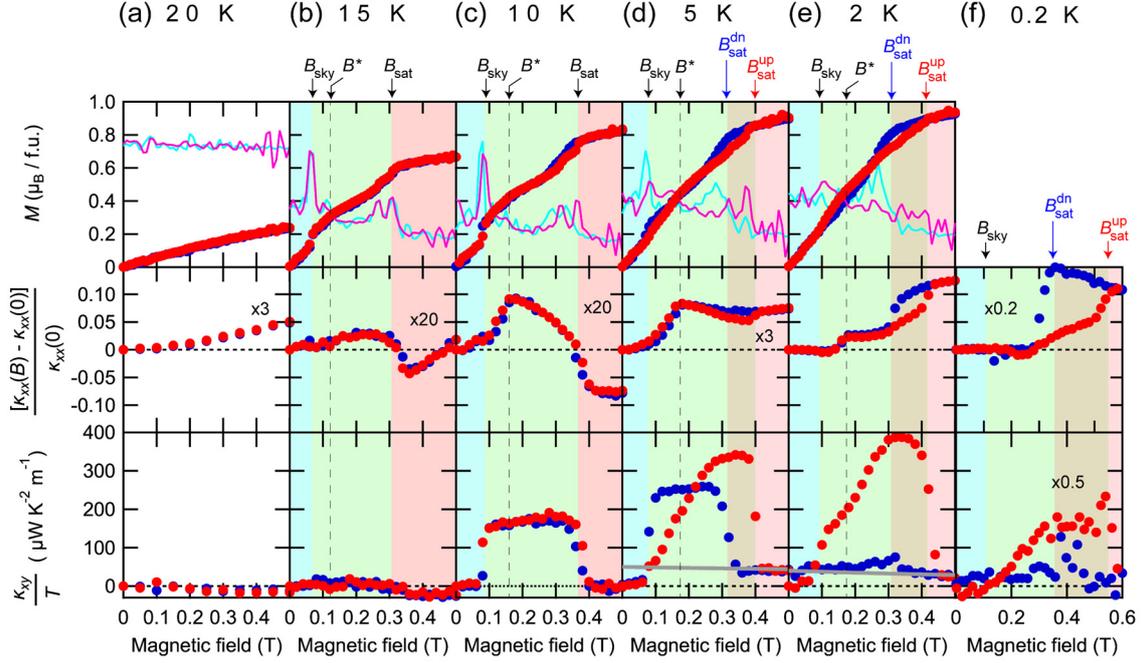

**Fig. 2 Thermal Hall effect by magnetic skyrmions.** Top, middle, and bottom panels show the field dependence of the magnetization ($M$), the field-induced deviation of the longitudinal thermal conductivity [$\kappa_{xx}(B) - \kappa_{xx}(0)$] normalized by the zero-field value, and the thermal Hall conductivity divided by the temperature ($\kappa_{xy}/T$), respectively. The field differential of the magnetization $\partial M/\partial B$ is shown in arbitrary unit in the top panels (solid lines). The data obtained in the magnetizing ("$|B|$ up") and that in the demagnetizing ("$|B|$ down") process are shown in red and blue, respectively. To show the magnetic hysteresis in the saturation field, $B_{\mathrm{sat}}$ in the $|B|$ up and that in the $|B|$ down process is marked separately at lower temperatures (d–f). For clarity, the data in the middle panels is multiplied by a constant indicated in the panel. The data of $\kappa_{xy}/T$ at 0.2 K is multiplied by 0.5 by the domain volume effect in the different cooling process (see S4 in SM). The grey solid lines in $\kappa_{xy}/T$ show an estimation of $\kappa_{xy}^{\mathrm{mag}}$.

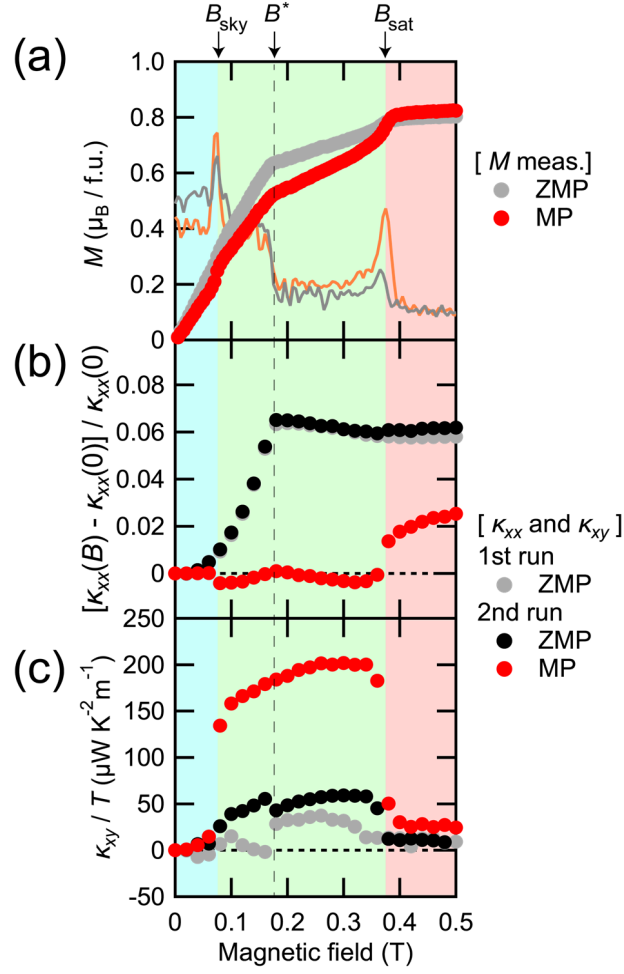

**Fig. 3 Magnetic-field poling effect.** (a–c), The field dependence of $M$ (a), $[\kappa_{xx}(B) - \kappa_{xx}(0)]/\kappa_{xx}(0)$ (b), and $\kappa_{xy}/T$ (c) at 5 K of sample 2. The field differential of the magnetization $\partial M/\partial B$ is shown in arbitrary unit as the solid lines. These measurements were done in the $|B|$ down process after cooling through $T_S$ under zero field (ZMP, grey and black symbols) and a finite field (MP, red symbols). See S5 in SM for other MP data at different temperatures.

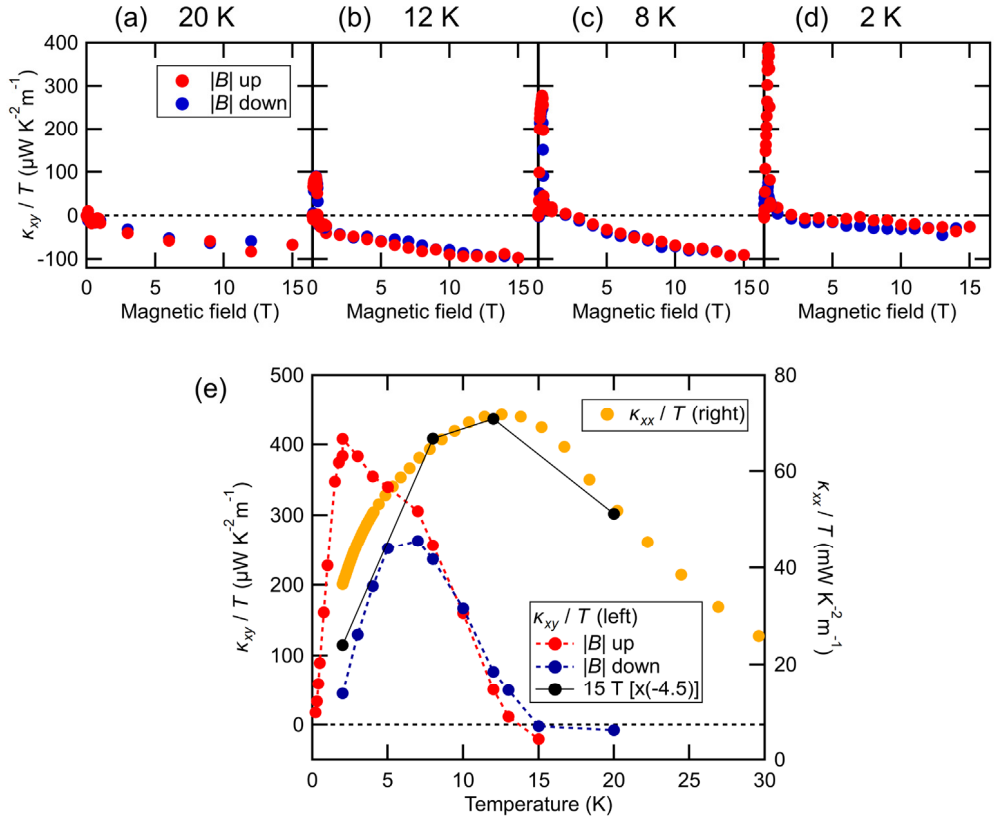

**Fig. 4 (a–d)** The field dependence of $\kappa_{xy}/T$ up to 15 T at 20 K (a), 12 K (b), 8 K (c), and 2 K (d). (e) The temperature dependence of $\kappa_{xy}/T$ (left) and that of $\kappa_{xx}/T$ (right, the same data shown in Fig. 1(e)). The black circles show $\kappa_{xy}/T$ at 15 T (from a–d) multiplied by $-4.5$, which well scales to $\kappa_{xx}/T$.

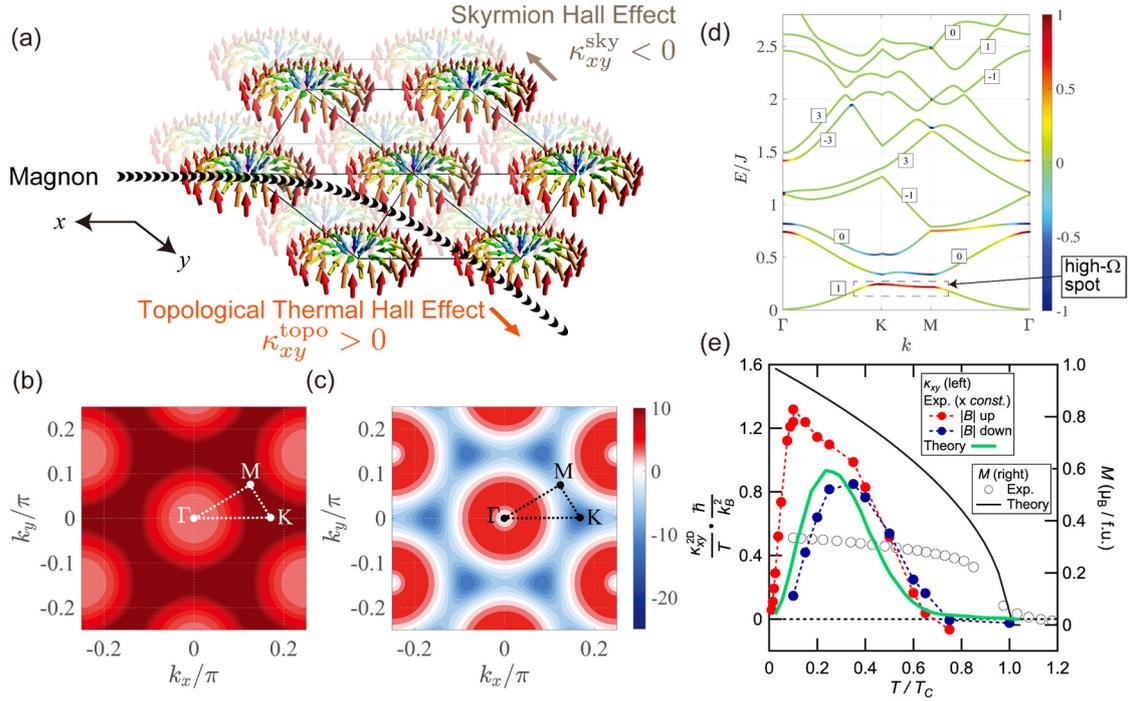

**Fig. 5 Theoretical calculations of topological Hall effects of magnons.** (a) a schematic illustration of the topological thermal Hall effects in a lattice of Néel-type skyrmions. Magnon thermal currents are bent by the Berry phase effect given by the skyrmion lattice, giving rise to the topological thermal Hall effect of magnons ($\kappa_{xy}^{\text{topo}} > 0$). The backaction of the topological thermal Hall effect produces the skyrmion Hall effect on the skyrmion lattice ($\kappa_{xy}^{\text{sky}} < 0$) [26]. (b, c) Color plots of the Berry curvature of the lowest (b) and the second lowest (c) energy bands of the magnons in a triangular lattice of the Néel-type magnetic skyrmions. (d) The energy bands along the high-symmetry points. The color denotes the Berry curvature that is normalized in each band. (e) The temperature dependence of $\kappa_{xy}^{\text{2D}}/T$ (left) and $M$ (right) obtained by the calculation (solid lines) and the experiments (symbols). The experimental data obtained at 0.36 T in the magnetizing ("$|B|$ up") and that at 0.2 T in the demagnetizing ("$|B|$ down") process are shown in red and blue circles, respectively. To fit with the theoretical result, the horizontal axis of the experimental $\kappa_{xy}^{\text{2D}}/T$ data is normalized by $T_\text{C} = 20$ K. The vertical axis of $\kappa_{xy}^{\text{2D}}/T$ in the VTI(DR) measurements are multiplied by 10(5) to further include the domain volume effect (see S4 in SM).

**Table 1** Full list of the Chern number $C_n$ of the magnon bands in the lattice of the magnetic skyrmions.

| $n$ | $C_n$ | $n$ | $C_n$ | $n$ | $C_n$ | $n$ | $C_n$ |
|---|---|---|---|---|---|---|---|
| 1 | 1 | 17 | 10 | 33 | 1 | 49 | -4 |
| 2 | 0 | 18 | 1 | 34 | 3 | 50 | 2 |
| 3 | 0 | 19 | -4 | 35 | -1 | 51 | 3 |
| 4 | -1 | 20 | 0 | 36 | -6 | 52 | 3 |
| 5 | 3 | 21 | -2 | 37 | 2 | 53 | 0 |
| 6 | -3 | 22 | 5 | 38 | 1 | 54 | -5 |
| 7 | 3 | 23 | 7 | 39 | -1 | 55 | 2 |
| 8 | -1 | 24 | -8 | 40 | -1 | 56 | 2 |
| 9 | 1 | 25 | 3 | 41 | 6 | 57 | -4 |
| 10 | 0 | 26 | -1 | 42 | -4 | 58 | 2 |
| 11 | -3 | 27 | 0 | 43 | -2 | 59 | 2 |
| 12 | 9 | 28 | 3 | 44 | 2 | 60 | 0 |
| 13 | -5 | 29 | -6 | 45 | 0 | 61 | 1 |
| 14 | -4 | 30 | 11 | 46 | -4 | 62 | -1 |
| 15 | 0 | 31 | -7 | 47 | -2 | 63 | 0 |
| 16 | -3 | 32 | -2 | 48 | -2 | 64 | -2 |

# Supplementary Materials for

## Topological Thermal Hall Effect of Magnons in Magnetic Skyrmion Lattice

Masatoshi Akazawa[1]†, Hyun-Yong Lee[2,3,4]†, Hikaru Takeda[1], Yuri Fujima[5], Yusuke Tokunaga[5], Taka-hisa Arima[5], Jung Hoon Han[6]*, Minoru Yamashita[1]*

Correspondence to: J.H.H. (hanjemme@gmail.com) and M.Y. (my@issp.u-tokyo.ac.jp)

**This Supplementary Materials includes:**

Supplementary Text
- S1. Magnetization data up to high magnetic fields
- S2. Raw data of the transverse temperature difference $\Delta T_y$
- S3. Sample dependence of the thermal Hall effect
- S4. Thermal-transport measurements of sample 1 in dilution refrigerator
- S5. Details of the magnetic-field poling measurements
- S6. Heat current dependence of the thermal Hall effect
- S7. Scattering effect on magnons by magnetic skyrmions

Figs. S1 to S11
References

## S1. Magnetization measurements up to high magnetic fields

The magnetization of GaV$_4$Se$_8$ was measured up to high fields as shown Fig. S1. The magnetization does not reach the full saturation ($1\mu_B$) even at saturation field $B_{\text{sat}}$ (~0.4 T), but asymptotically approaches to the full saturation as increasing the magnetic field. This gradual increase of $M$ above $B_{\text{sat}}$ is caused by the multi-domain effect and the magnetic easy-plane anisotropy [1]. As discussed in the main text, the structural transition at $T_S = 41$ K results in four crystallographic polar domains elongating to either [111], [$\bar{1}$11], [1$\bar{1}$1], and [11$\bar{1}$]. Although $M$ in the domain with the elongating axis parallel to $B$ fully contributes the measured magnetization, those in other three domains are tilted from $B$ because of the magnetic uniaxial anisotropy along the elongating axis, resulting in the gradual increase of $M$ as shown in Fig. S1. A similar field dependence of the magnetization has been measured in GaV$_4$S$_8$ (Ref. [2,3]).

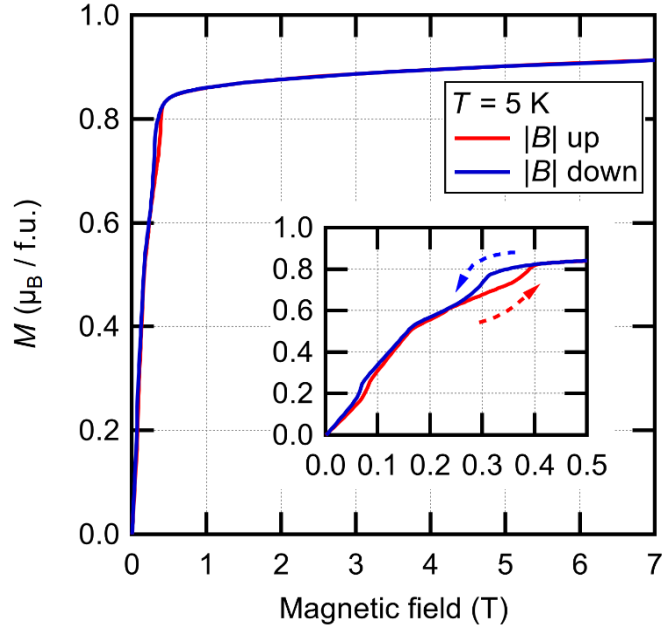

**Fig. S1** The field dependence of the magnetization up to high field at 5 K. The magnetic field was applied along [111] of the sample. The inset shows an enlarged view of the low field data.

## S2. Raw data of the transverse temperature difference $\Delta T_y$

Here we explain the procedure of the thermal Hall measurement to obtain $\Delta T_y^{asym}$ in Eq. S1.

A typical field dependence of $\Delta T_y$ at a fixed temperature is shown in Fig. S2b. To investigate a difference of $\kappa_{xy}$ by entering the skyrmion phase from the low-field cycloidal phase and from the high-field forced-ferromagnetic phase, we measured the field dependence of $\Delta T_y$ both in the magnetization ("$|B|$ up") and demagnetization ("$|B|$ down") process under the positive and the negative fields (Fig. S2a). We then antisymmetrized $\Delta T_y$ with respect to the field direction for each magnetizing (red and orange data in Fig. S2b) and the demagnetizing (blue and cyan data in Fig. S2b) process to obtain $\Delta T_y^{asym}$ for both procedures (Fig. S2c). We confirmed that there is no discernible difference in the $|B|$ up measurements done after the zero-field cooling through $T_C$ and those after the $|B|$ down measurements. We also confirmed that a $|B|$ up measurement after a field cooling at $B_{sky} < |B| < B_{sat}$ shows the same result with that after the zero-field cooling.

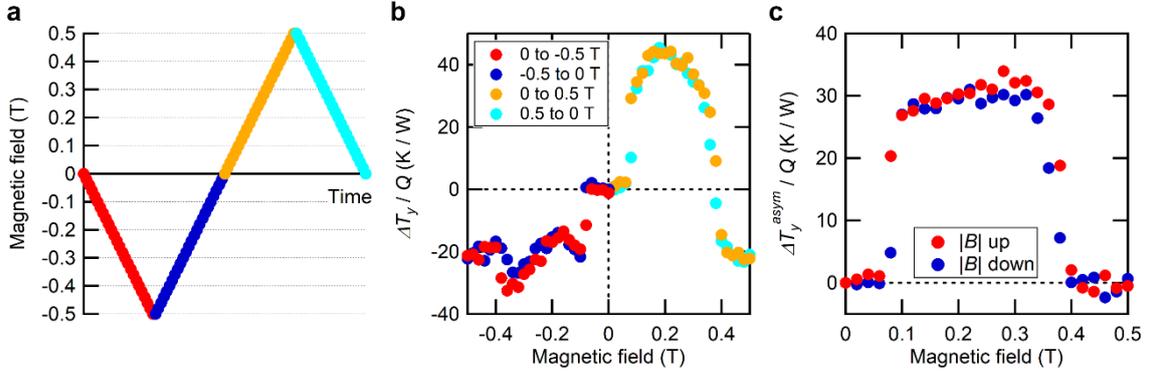

**Fig. S2** The field dependence of the transverse temperature difference $\Delta T_y = T_{L1} - T_{L2}$. **a**, The field procedure of the thermal Hall measurement. **b**, The field dependence of $\Delta T_y/Q$ of sample 1 observed at 10 K. **c**, The field dependence of $\Delta T_y^{asym}/Q$ obtained for magnetizing (red) and demagnetizing (blue) process.

**S3. Sample dependence of the thermal Hall effect**

Figures S3 to S6 show the data of sample 2. We confirmed the reproducibility of all the data of sample 1 whereas the magnitude of $\kappa_{xy}$ of sample 2 is smaller than that of sample 1 (Fig. S6). This small thermal Hall signal is attributed to a smaller fraction of the domain volume of the magnetic skyrmion phase. This smaller fraction of the skyrmion domain volume can be seen in the smaller increase of $M$ at $B_{sky}$ of sample 2 (top panels in Fig. S4) than that in sample 1 (see Fig. 2 in the main text). The relation between the skyrmion domain volume and $\kappa_{xy}$ is confirmed by the magnetic-field poling, as discussed in the main text and section S5.

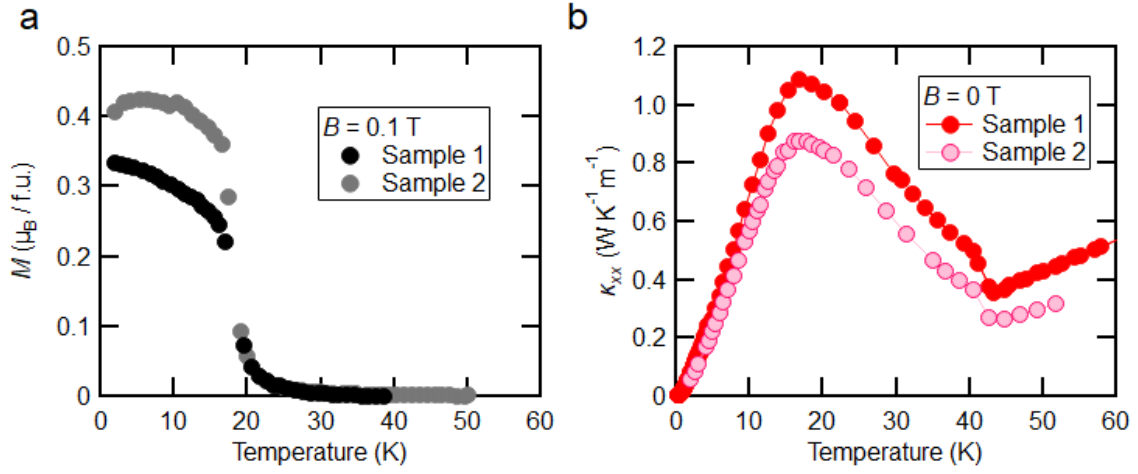

**Fig. S3** The temperature dependence of the magnetization ($M$, **a**) and the longitudinal thermal conductivity ($\kappa_{xx}$, **b**) of sample 1 and sample 2.

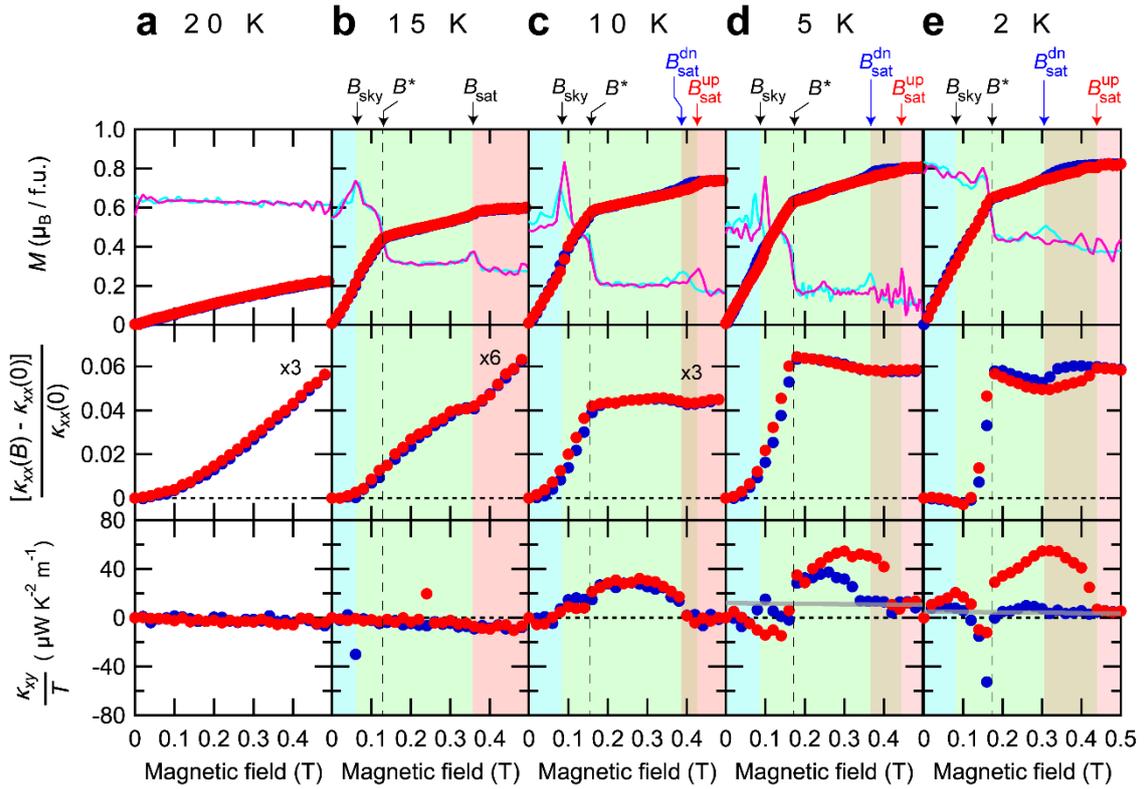

**Fig. S4 a-e,** The field dependence of the magnetization ($M$, top panels), the field-induced deviation of the longitudinal thermal conductivity $[\kappa_{xx}(B) - \kappa_{xx}(0)]$ normalized by the zero-field value $[\kappa_{xx}(0)]$ (middle panels), and the thermal Hall conductivity divided by the temperature ($\kappa_{xy}/T$, bottom panels) at different temperatures of sample 2. The field differential of the magnetization $\partial M/\partial B$ is shown in arbitrary unit in the top panels (solid lines). The data obtained in the magnetizing ("$|B|$ up") and that in the demagnetizing ("$|B|$ down") process are shown in red and blue, respectively. For clarity, the data in the middle panels is multiplied by a constant indicated in the panel. The grey solid lines in $\kappa_{xy}/T$ show an estimation of $\kappa_{xy}^{\mathrm{mag}}$ (see section S6).

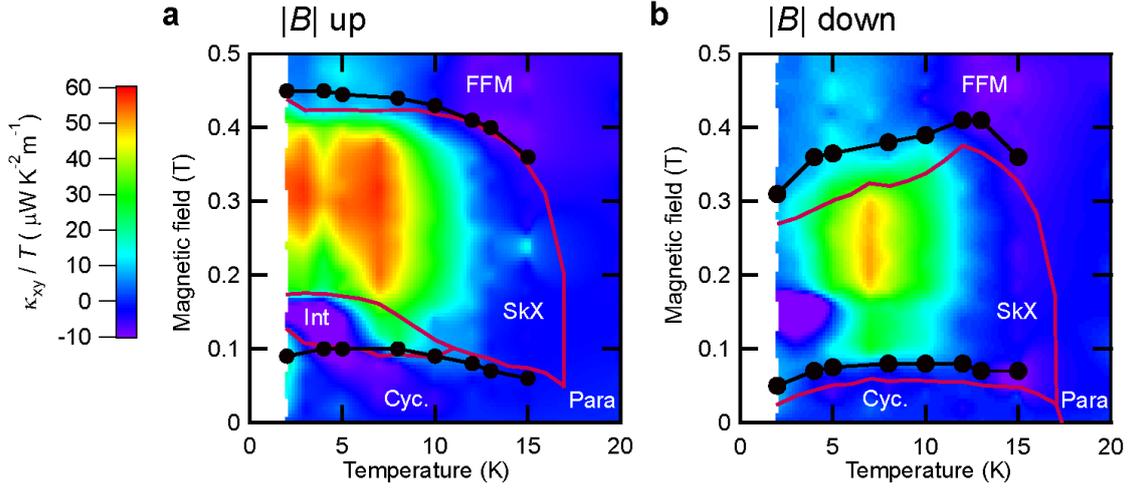

**Fig. S5 a,b** Color plots of the thermal Hall conductivity divided by the temperature ($\kappa_{xy}/T$) of sample 2 measured in the magnetizing ($|B|$ up, **a**) and demagnetizing ($|B|$ down, **b**) procedure. The red solid lines show the magnetic phase boundary determined by the previous measurement [4]. The black circles show the phase boundary of sample 2 determined by the magnetization measurement (top panels of Fig. S4).

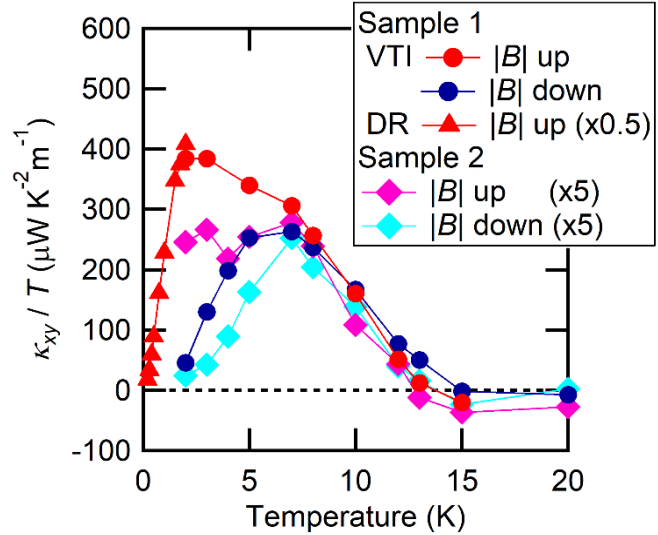

**Fig. S6** The temperature dependence of $\kappa_{xy}/T$ at 0.36 T ($|B|$ up) and that at 0.2 T ($|B|$ down, blue circles) of sample 1 and sample 2. The dilution refrigerator (DR) data of sample 1 is multiplied by 0.5 as described in the next section. The data of sample 2 is multiplied by 5 for a comparison.

## S4. Thermal-transport measurements of sample 1 in dilution refrigerator

To investigate a possible magnetic transition in the magnetic skyrmion phase below 2 K, the thermal-transport measurements of sample 1 were extended down to 0.2 K by using a dilution refrigerator (DR) after the VTI measurements. A change of the magnetic structure from the magnetic skyrmion phase to another magnetic phase must accompany changes both in $\kappa_{xx}$ and in $\kappa_{xy}$. The former is caused by a change in the scattering of phonons or magnon by the magnetic transition. The latter is brought by a change in the Berry phase distribution in the magnon band structure. However, as shown in Fig. S7, no discernible anomaly is seen in the temperature dependence of $\kappa_{xx}$ in the skyrmion phase, showing the absence of another magnetic phase transition inside the magnetic skyrmion phase. On the other hand, at 0.5 T ($B > B_{\text{sat}}$), a hump-like increase is observed at the phase boundary between the forced-ferromagnetic and the magnetic skyrmion phase, showing the sensitivity of $\kappa_{xx}$ to a magnetic transition. This increase also shows the magnon contribution in $\kappa_{xx}$. Moreover, as shown in Fig. S8, the field dependence of $\kappa_{xy}/T$ smoothly varies with smaller magnitude as lowering temperature below 2 K, further evidencing the absence of another magnetic phase transition.

As shown in Fig. S8a, the field dependence of $\kappa_{xy}/T$ at 2 K is almost twice as large as that in the VTI at the same temperature, showing that the fraction of the skyrmion domain volume is doubled in this cooling process. On the other hand, $\kappa_{xx}$ shows discernible difference between the two measurements (see Fig. 1E in the main text), indicating that the skyrmion domain volume only affects $\kappa_{xy}$ (see also Fig. S9a). To take into account this domain size effect on $\kappa_{xy}$, all the DR data of sample 1 shown in the main text is multiplied by 0.5. We then multiply a constant to the experimental data shown in Fig. 5e, to fit with the theoretical result. Note that, data of the DR measurements in the demagnetization process are not shown in Fig. 5e, because the appearance of the skyrmion signal is no longer discernible in the field dependence of $\kappa_{xy}$ below 2 K as shown in Fig. S8.

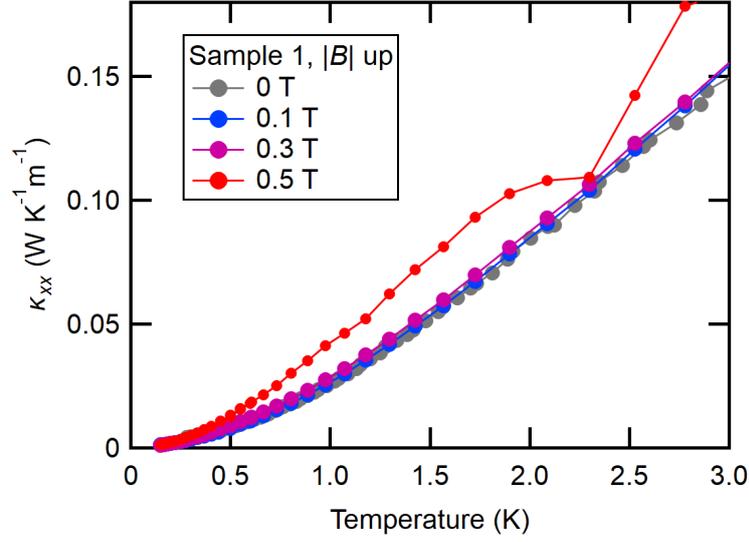

**Fig. S7** The temperature dependence of $\kappa_{xx}$ of sample 1 at different magnetic fields. The data was measured in the $|B|$ up process in the DR.

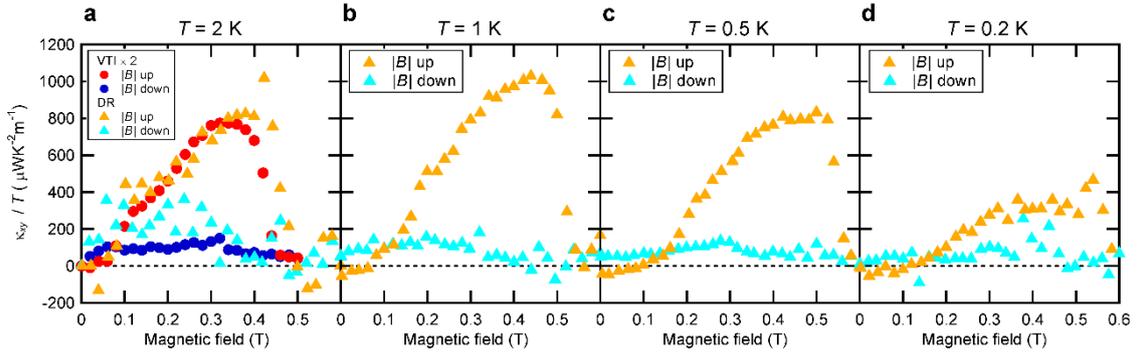

**Fig. S8 a-d**, The field dependence of $\kappa_{xy}/T$ of sample 1 in the DR measurements (triangles). The magnitude of $\kappa_{xy}/T$ is almost twice as large as that observed in the previous VTI measurements (circles).

## S5. Details of the magnetic-field poling measurements

Here, we show details of the magnetic-field poling (MP) measurements. Prior to the thermal-transport measurements, we performed the magnetization measurements of sample 2 by cooling under zero field (ZMP), which was followed by the subsequent measurements cooled under 7 T at the structural transition (MP). As shown in Fig. 3A in the main text, we confirm that the MP increases the skyrmion domain volume. After the 1st run measurements of the thermal conductivity of sample 2 with the zero field at $T_S$ (denoted as "1st run ZMP"), we cooled down sample 2 again at zero field ("2nd run ZMP"), warmed up above $T_S$ and cooled down with 15 T when passing through $T_S$ ("2nd run MP"). We find that, although this MP does not affect $\kappa_{xx}$ at zero field (Fig. S9a), both the field dependence of $\kappa_{xx}$ and that of $\kappa_{xy}$ are drastically changed by MP (Fig. S9b and S10).

First, this MP enhances the magnitude of $\kappa_{xy}$. As shown in Fig. S10, whereas the field dependence of $\kappa_{xy}/T$ in the ZMP process is similar both in 1st and 2nd run, $\kappa_{xy}/T$ in the MP process becomes about three times larger than that in the ZMP, demonstrating that the magnitude of $\kappa_{xy}/T$ reflects the skyrmion domain volume increased by the MP.

Second, the gradual increase of $\kappa_{xx}$ by magnetic field is taken over by a drop of $\kappa_{xx}$ as entering the magnetic skyrmion phase. As clearly seen 10 or 5 K data in Fig. S10 (see also Fig. 3 in the main text), $\kappa_{xx}$ in the ZMP process gradually increases to $B^*$, which is followed by a saturation above $B^*$. This gradual increase almost disappears by the MP. Instead, a sharp decrease is observed at the boundary between the cycloidal and the magnetic skyrmion phases. This field suppression effect in the magnetic skyrmion phase can be caused by either the backflows of the skyrmion motion [5,6] or additional scattering effects by skyrmions on magnons and phonons. Although the thermal diffusion of the magnetic skyrmions from the hot to the cold region has been observed in a ferromagnetic metal film at a very high temperature [7], such a positive contribution from the skyrmion flow is not observed in our experiments. The gradual increase of $\kappa_{xx}(B)$ in the ZMP is brought by the increase of the phonon conductivity in the other domains, in which the conical phase asymptotically approaches to the forced-ferromagnetic phase.

Finally, this MP effects reveal the decrease of $\kappa_{xx}$ above $B_{\text{sat}}$ for 5 K $< T <$ 20 K, which is masked by the increase of $\kappa_{xx}$ brought by the conical phase in the other domains in the ZMP measurements. In the forced-ferromagnetic phase, there are contrasting field-induced effects on the thermal conductions of magnons ($\kappa_{xx}^{\text{mag}}$) and phonons ($\kappa_{xx}^{\text{ph}}$). The magnon population is suppressed in the forced-ferromagnetic phase by the field-induced gap. Therefore, $\kappa_{xx}^{\text{mag}}$ is decreased under a magnetic field. On the other hand, $\kappa_{xx}^{\text{ph}}$ is increased because magnon-phonon scattering is suppressed by the spin polarization. Therefore, the negative magneto-thermal conductivity observed in the forced-ferromagnetic phase at 10 and 15 K indicates the decrease of $\kappa_{xx}^{\text{mag}}$ by the field-induced gap, which in turn shows the presence of $\kappa_{xx}^{\text{mag}}$ in the skyrmion phase. For $T \leq$ 5 K, this negative magneto-thermal conductivity is taken over by the increase of $\kappa_{xx}^{\text{ph}}$ by the suppression of the magnon-phonon scattering.

All these MP effects on $M$, $\kappa_{xx}$, and $\kappa_{xy}$ demonstrate the increased volume of the magnetic skyrmion domain and genuine features of the magnetic skyrmion phase. We find that these MP effects were stable during the whole series of measurements (which typically takes 1–2 months) as long as the sample was kept below $T_S$.

The temperature dependence of $\kappa_{xy}/T$ in the $|B|$ up (0.36 T, triangles) and that in the $|B|$ down (0.2 T, reversed triangles) process is summarized in Fig. S9b. In addition to the enhancement of the magnitude of $\kappa_{xy}/T$, we find that $\kappa_{xy}/T$ in the $|B|$ down process below 5 K is increased in the 2nd run even in the ZMP. As discussed in the main text, we suggest that a subtle rearrangement in the skyrmion lattice causes the large hysteresis in $\kappa_{xy}/T$ at lower temperatures. The difference of $\kappa_{xy}/T$ in the ZMP of the 1st and the 2nd run below 5 K might also be related to this rearrangement.

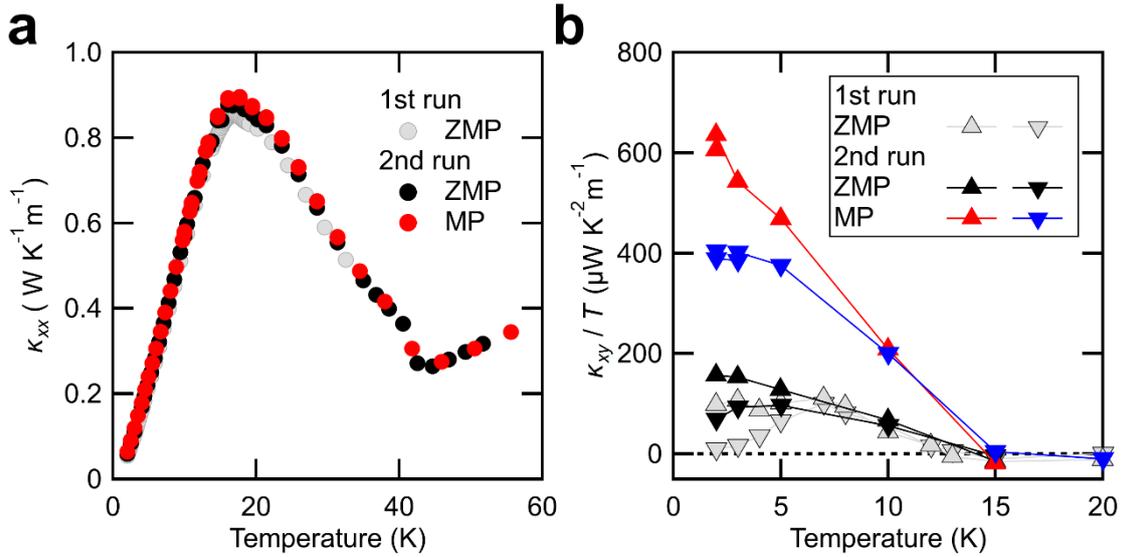

**Fig. S9 a**, The temperature dependence of $\kappa_{xx}$ of sample 2 in the ZMP and MP measurements. **b**, The temperature dependence of $\kappa_{xy}/T$ in the $|B|$ up (0.36 T, triangles) and that in the $|B|$ down (0.2 T, reversed triangles) process. These measurements were done in sample 2 after cooling through $T_S$ under zero field in the 1st and the 2nd run (ZMP, grey and black symbols) and a finite field the 2nd run (MP, red symbols).

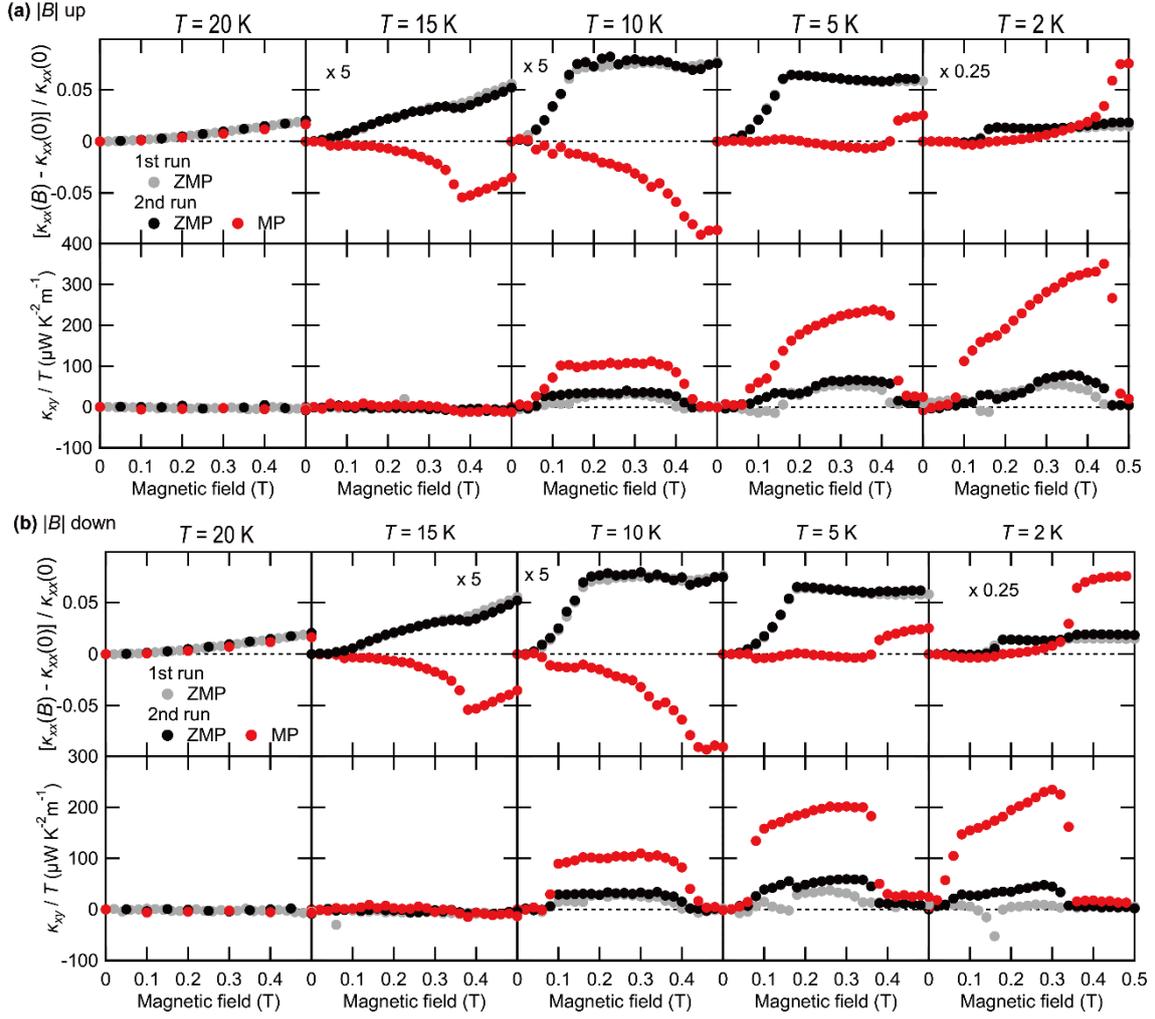

**Fig. S10 a, b** Field dependence of the field-induced deviation of the longitudinal thermal conductivity $[\kappa_{xx}(B) - \kappa_{xx}(0)]$ normalized by the zero-field value (top panels), and the thermal Hall conductivity divided by the temperature ($\kappa_{xy}/T$, bottom panels) observed in the magnetization ("$|B|$ up", **a**) and the demagnetization ("$|B|$ down", **b**) measurements of sample 2. The data obtained in the 1st run ZMP, 2nd run ZMP, and 2nd run MP is shown by grey, black and red circles, respectively.

## S6. Heat current dependence of the thermal Hall effect

To investigate the depinning threshold of the magnetic skyrmions in GaV$_4$Se$_8$, we checked the heat current $Q$ dependence by applying $Q$ as large as possible at several temperatures in the magnetic skyrmion phase. As shown in Fig. S11, both the longitudinal ($\Delta T_x$) and the transverse ($\Delta T_y^{asym}$) temperature difference show the linear $Q$ dependence in the whole $Q$ range we investigated, demonstrating that both $\kappa_{xx}$ and $\kappa_{xy}$ do not depend on $Q$. This absence of the $Q$ dependence in $\kappa_{xy}$ indicates either the depinning threshold is too large or too small to observe. In the former case, there is no skyrmion Hall effect because the magnetic skyrmions are completely pinned by a strong pinning potential. In the latter, the magnetic skyrmions are flowing with a constant friction kept in the whole $Q$ range of our measurements, giving rise to a huge reduction of $\kappa_{xy}^{topo}$ in our measurements. Although the latter scenario is rather speculative, this skyrmion Hall effect might explain the reduction of the experimental $\kappa_{xy}$ data compared to the theoretical estimation, in addition to the domain volume effect. In either case, however, we can safely conclude that the thermal Hall effect is dominated by $\kappa_{xy}^{topo}$.

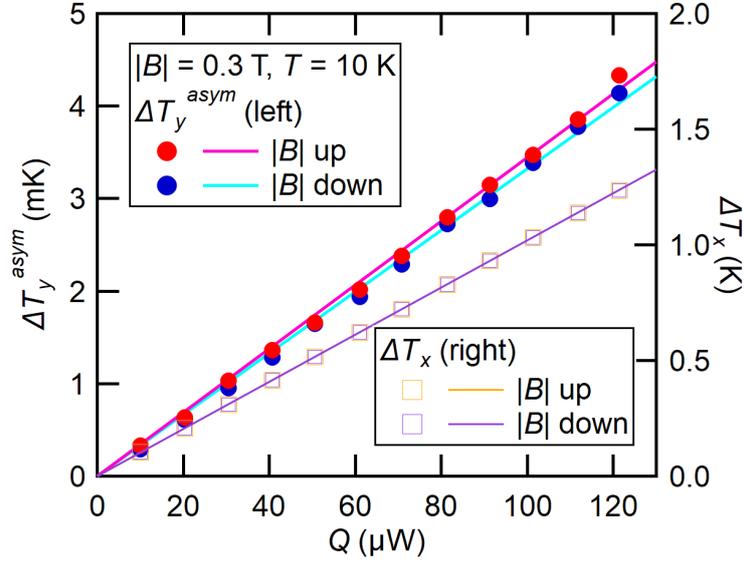

**Fig. S11** The thermal current ($Q$) dependence of the longitudinal ($\Delta T_x$, right axis) and the transverse ($\Delta T_y^{asym}$, left axis) temperature difference of sample 1 at 10 K. Note that $\Delta T_x$ data of $|B|$ up and that of $|B|$ down are overlapped each other.

### S7. Scattering effect on magnons by magnetic skyrmions

Here, we discuss an order estimation about the scattering effects on magnons by an individual magnetic skyrmion. The emergent field produced by a skyrmion is given by

$$B_{em} = -\frac{h}{e}\frac{2}{\sqrt{3}\, a_{sk}^2},$$

where $a_{sk}$ is the lattice constant of the skyrmion lattice. According to the neutron experiment [2], $a_{sk}$ is estimated as 22.2 nm, which results in $B_{em} = 9.5$ T. As shown by Ref. [6], this emergent field works to scatter the magnon, giving rise to a peak of the magnon-skyrmion scattering angle at $k\xi \sim 1$, where $k$ is the wave number of the incoming magnon and $\xi$ is the core size of the skyrmion. The Debye distribution of bosons in two dimensions $(x^3 e^x)/(e^x - 1)^2$, where $x = \hbar\omega/k_B T$, gives the maximum of the magnon population at $\hbar\omega = J(ka)^2 \sim 2.6 k_B T$, where $J$ is the magnetic interaction energy and $a \sim 7$ Å is the size of the V$_4$Se$_4$ unit cell. Therefore, the wave number of the maximum population becomes smaller (wave length becomes larger as expected for bosons) at lower temperatures, giving rise to the peak in the scattering angle when $k \sim 1/\xi$. Since the core size of the skyrmion is estimated as $\xi \sim 10a$, the Hall angle has a peak at $J\left(\frac{a}{10a}\right)^2 \sim 2.6\, k_B T$, or

$$\frac{k_B T}{J} \sim \frac{1}{260}.$$

Given that $\kappa_{xy}$ is given by the product of the Hall angle and the magnon thermal conductivity $\kappa_{xx}^{\text{mag}} \sim T^\alpha$ ($\alpha = 2$–$3$), the peak temperature of $\kappa_{xy}/T$ could be much lower than this estimation. Therefore, this scattering picture is inconsistent with our experimental data of GaV$_4$Se$_8$ showing the peak of $\kappa_{xy}/T$ at $k_B T/J \sim 1/3$ (see Fig. 4e).

In contrast, the Berry phase effect in the reconstructed magnon band brought by our momentum-based picture adopted in the main text works as long as the skyrmion lattice exists. Further, our momentum-based picture has a clear benefit allowing one to obtain $\kappa_{xy}$ directly from the Berry curvature distribution through the relation (Eq. (1))

$$\frac{\kappa_{xy}}{T} = \frac{k_B^2}{\hbar} \int \Omega(E) f(E) dE.$$

In fact, our calculation well reproduces the temperature dependence of our $\kappa_{xy}$ data in the $|B|$ down process (see Fig. 5 in the main text) with very little material input parameters. In contrast, as we discussed above, the scattering picture requires the knowledge of the longitudinal thermal conductivity of magnons $\kappa_{xx}^{\text{mag}}$ (i.e. details of the scattering time, velocity, etc. of magnons) which is unattainable for most magnetic insulators because both phonon and magnons contribute to the total $\kappa_{xx}$ in a complex manner.